# RECOMMENDATION SYSTEM FOR INFORMATION SERVICES ADAPTED, OVER TERRESTRIAL DIGITAL TELEVISION

Mery Y. Uribe Rios[1] and Rafael V. Páez[2]

[1] mery.uribe@javeriana.edu.co
[2] paez-r@javeriana.edu.co

[1,2] Pontificia Universidad Javeriana, Bogotá, Colombia

## ABSTRACT

*The development of digital television in Colombia has grown in last year's, specially the digital terrestrial television (DTT), which is an essential part to the projects of National Minister of ICT, thanks to the big distribution and use of the television network and Internet in the country. This article explains how joining different technologies like social networks, information adaptation and DTT, to get an application that offers information services to users, based on their data, preferences, inclinations, use and interaction with others users and groups inside the network.*

## 1. INTRODUCTION

The development of apps for digital television in Colombia is a starting process, due to the state of this technology in the country and to the projects of Ministry of ICT to the expansion of DTT, with the aim of replace the analogical television, which currently is over the most part of the national territory [1] [2]. Because of this, the implementation of applications of this type represents an opportunity of innovation, participation and research in this context.

Being a part of this innovation, we must participate with original ideas that would bring benefits for the finals users and for the designers, and developers of those apps. We are trying to enclose these issues with the outgoing interaction user data in the network in DTT, in order to adapt services according to the specific needs of each user. All this, to take advantage of the social network boom, the benefits of services adapted and the big distributed network deployed for all users of DTT.

This document explains the design, development and implementation of a social network for DTT used as a recommendation system, based on the interaction among its users.

## 2. SOCIAL NETWORK FOR INFORMATION SERVICES OVER DTT

### 2.1. Principal Base Architecture

The interactive app for DTT is part of a project called "Architecture of Adaptation of Information Services over Digital Television, supported on a social network". With the implementation of this app, we want to validate some aspects of this architectural design and make of this a study case.
The architectural styles to validate through the application are: separation of the composition, design and logic of an app, client – server, architecture for layers, the use of a demand service, separation of files for server and separation of services.

The validation was made through the implementation of the app over an environment of digital terrestrial television, with a publicity recommendation system based on the interaction of television users in the social network.

### 2.2. Adaptation of information

Based on system requirements and some services, which were taken as a study case to materialize and exemplify a possible service for offering in the architecture, such as *i)* To offer a publicity recommendation and *ii)* To notify a recommendation's state, we defined a model of adaptation based on profiles for enhance services due to their specification and to the needs of users, through the management of some determinate data. The use of profiles lets to the user or group data process , locating and using specific information for each case, as is mentioned in [3] "all the get data must be integrated to create complete profiles of the consumers". In this way, we can represent preferences in a orderly, automatic way [4–7] and we can to find relevant information which can to be used through processes like user profile mining [8] or recommendations services [9]. It should be noted that this adaptation model is not limited to the services mentioned, they are just a study framework; the model is general and transversal to any service.

We worked with five types of profiles, according to the kind of information to manage in each one. The profiles were:

1. User profile: Stores information about basic data and user preferences.
2. Context profile: Uses information about events that involve the interaction of user and the system [10]. Those data were needed to monitor some variables that can influence, change and/or improve the services offered by the system. Also, these data were taken from the television's information system and the monitoring the system itself.
3. Device profile: Specifies the information related to the access device. These data were taken in count in order to adapt the information and services, according to the respective features of each device, particularly in some ones with limited capabilities, as mobile devices. These data will be taken from CC/PP file CC/PP [11] of each device.
4. Social profile: Specifies the interactions of a user into the social network. These data describe how a user is related to others users, group(s), services provided by the system and with information about television content. All these behaviors are the most common patterns of interaction recognized. Those data were taken from the interactions of each user into the social network.
5. Group profile: Describes the data of a group of users, it can be generated by the system or created by the user. We want to store management data about the group and its different relationships with services and users. In this way is possible to work with each preferences' group in the same way that the user's ones. In this way, the services can be guided to a set of users with some features according to specific preferences of each group. These data will depend on the way them were created in the app (for the system or a user).

Computer Science & Engineering: An International Journal (CSEIJ), Vol.2, No.6, December 2012

The adaptation model will be given for the follow figure:

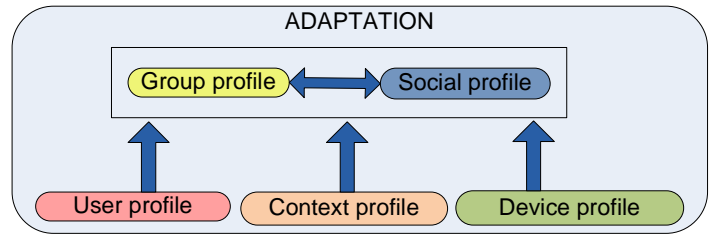

Figure 1. Adaptation model

This figure shows the dependence of group and social profiles with the user's one, and context profiles to relate to the users. The creation and generation of groups and the interactions among users will depend on the correlation between data of the profiles to find the necessary information to offer services. It should be noted that the services to adapt not only depend on the social and group profiles, but also to link information from all profiles.

### 2.3. Social Network

Some aspects like interactions among users, nodes network distribution, its representation and how to analyze its data are taken in account for the network design.

According to the options of network data management explained in [12], is recommendable to represent the network data like a graph, because it allows handling them in a generally and characteristically than a sociogram (representation of a social environment, where the nodes or points represent people, their goals or actions, while the connections represent the sequence that connects them [12]).

Each interaction of a user (node) into the network will be represented like an arc in the graph, depending on its type and its correlation with other data. These nodes are characterized by two aspects: user and access device. In the last time, and depending on the reason of the relationship among nodes, there are two kinds of interactions:

- Interaction of use: Related to the data profiles of each user and his activities into the network. These interactions represent relationship between user – user, user – group and user – service.

- Interaction of resource: Related to technical aspects of the access device. These data will be used to create interactions of resources, according to the technical capabilities and resources that each node needs to deploy a service.

In order to identify groups and opportunities to offer services, central points in the network are defined, which are points with the highest degree (highest number of arcs related (interactions)). This analysis allows to create an interaction among nodes, for instance, an interaction among users or among nodes for resources.

For finish this part, the distribution services method was defined, which will be "snow ball" technique. This technique allows using the relationships among users like a "clinch" to nominate to other users to a service or to nominate a service to a user [12]

## 2.4. Technical Aspects

To the implementation of an interactive app for digital television, we have taken into account some technical issues, especially those related to the digital television standards DVB-T y DVB-MHP [13–15], because they frame the activity in this context.

In this section the principal issues about the creation (design and implementation) used for the app is presented.

### 2.4.1. User interface

One of the relevant aspects in the design and development of an interactive app for digital television is the user interface, because the access devices and the execution environment are different compared with an app for computer or mobile device's one. The issues more outstanding are those related to the navigability and the design patterns.

#### 2.4.1.1. Navigability

This aspect frames the principal tips of the look and order of an app. The app control is specified through the use of navigation's buttons of a remote control, to manage the form and content of a menu.

In first place, the remote control (see Figure 2) is the device by which the user interacts with the television set and therefore, with the apps deployed in it. These remote controls have, generally, four colors buttons (red, green, yellow and blue), that together with navigation arrows, are used as controls for an application.

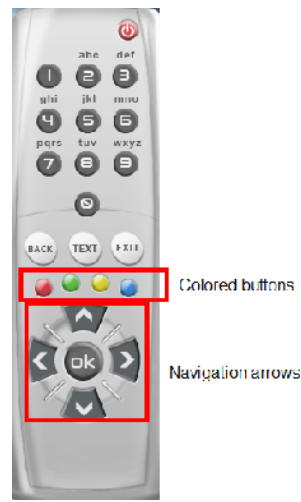

Figure 2. Remote control of TV set

Generally, each color key is associated with a determinate action, owing to the relationships of the user with the different colors [16], the
Figure 3 shows the principal actions for each color.

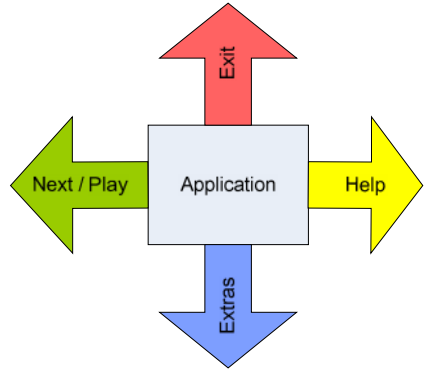

Figure 3. Actions for colors [16]

In second place, is important to be carefully managing the menu content and navigation; the information must be clear and concise, there should be no excessive submenus and should be easy to move back and forth through them.

So, we reach an important point for the user interface of this type of app, the access device. As this device can be a TV set, a mobile device or a computer, we can't extend the content or the navigation map. Although, we use the adaptation for taking in account some technical features to the access device, we have to use a little more for content adaptation, so the app must be designed to be deployed in any device.

### 2.4.1.2. Designed patterns

According to the application kind and the information amount to handle, the principal aspects can have different behaviors. These aspects are:

- Application location [16] (see Table 1)

Table 1. Location aspects

| Location | Amount of information displayed | Size | Location in the screen |
|---|---|---|---|
| **Superposition** | Little information related to content showed in the moment. | As small as possible. | Bottom or top of the screen.<br><br>No cover logos. |
| **Full screen with video** | So much content | Use ¼ of size screen for video in the most applications. | The video must be located on left top of the screen. |
| **Full screen without video** | So much content.<br><br>Use of complex functions not related to the TV program. | All screen | All screen |

4Computer Science & Engineering: An International Journal (CSEIJ), Vol.2, No.6, December 2012

- Menu [16]: to provide access to content and functionality (see Table 2).

Table 2. Menu aspects

| Location | Information deployed | Use of content | Highlight of content |
|---|---|---|---|
| **Simple menu** | Use 3 to 8 menu options | Select short and accurate titles for each menu | Provide a preview of the option, for instance a text or an image.<br><br>Mark the selected element. |
| **Multi screen** | Use for app that provides multiple video sources.<br><br>Use 3 to 8 videos sources. | Play the sound of the video source selected | Mark the selected element. |
| **Index** | Select a structure like an index that allows to access in a quickly way to popular elements | The application contents, some contents and options | Mark the selected element. |
| **Tabs** | Use them just for users who work on Internet. | Use horizontal tags. | Select short and accurate words for the title of each tab. |

**2.4.2. Design and implementation**

Having the mentioned aspects and specs on the system architecture, the user interface of the app was designed and deployed. Some tools to do this process were used. Icareus iTV Suite 3.9 [17] was used for the graphic interface, which "is an intuitive and versatile visual content development tool which enables rapid development of digital TV content" [17]. This app allows to the user interface to integrate external components, which are building by a software programmer, to give new functionalities to an app , according to the needs. The principal activities were:

**2.4.2.1. Design**

The main displays (see Figure 4) of app was designed according to the topics about designed patterns and having in account the important information to show in each one, because, it is an application which is deployment when a TV content is transmitted.

As this application is highly interactive, different aids were used, which helped to this interaction to be easier for using. For instance, some dialogs or a virtual keyboard were deployed, because the remote control's numeric keyboard of the TV set is not comfortable for the text exchange (see Figure 5 and Figure 6).

Computer Science & Engineering: An International Journal (CSEIJ), Vol.2, No.6, December 2012

- Menu [16]: to provide access to content and functionality (see Table 2).

Table 2. Menu aspects

| Location | Information deployed | Use of content | Highlight of content |
|---|---|---|---|
| **Simple menu** | Use 3 to 8 menu options | Select short and accurate titles for each menu | Provide a preview of the option, for instance a text or an image.<br><br>Mark the selected element. |
| **Multi screen** | Use for app that provides multiple video sources.<br><br>Use 3 to 8 videos sources. | Play the sound of the video source selected | Mark the selected element. |
| **Index** | Select a structure like an index that allows to access in a quickly way to popular elements | The application contents, some contents and options | Mark the selected element. |
| **Tabs** | Use them just for users who work on Internet. | Use horizontal tags. | Select short and accurate words for the title of each tab. |

**2.4.2. Design and implementation**

Having the mentioned aspects and specs on the system architecture, the user interface of the app was designed and deployed. Some tools to do this process were used. Icareus iTV Suite 3.9 [17] was used for the graphic interface, which "is an intuitive and versatile visual content development tool which enables rapid development of digital TV content" [17]. This app allows to the user interface to integrate external components, which are building by a software programmer, to give new functionalities to an app , according to the needs. The principal activities were:

**2.4.2.1. Design**

The main displays (see Figure 4) of app was designed according to the topics about designed patterns and having in account the important information to show in each one, because, it is an application which is deployment when a TV content is transmitted.

As this application is highly interactive, different aids were used, which helped to this interaction to be easier for using. For instance, some dialogs or a virtual keyboard were deployed, because the remote control's numeric keyboard of the TV set is not comfortable for the text exchange (see Figure 5 and Figure 6).

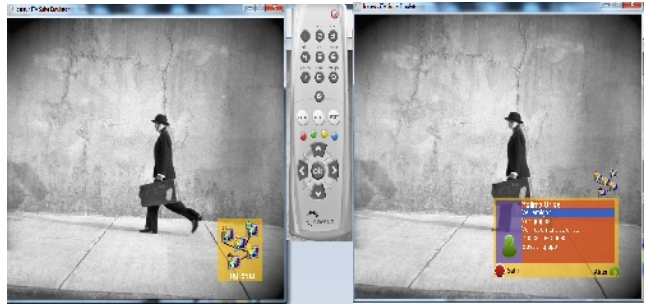

Figure 4. User interface

Throughout the application we can see the designed patterns mentioned, for instance, the use of the actions of each buttons color (see Figure 7), or the preview of the menu and sub menu options (see Figure 8).

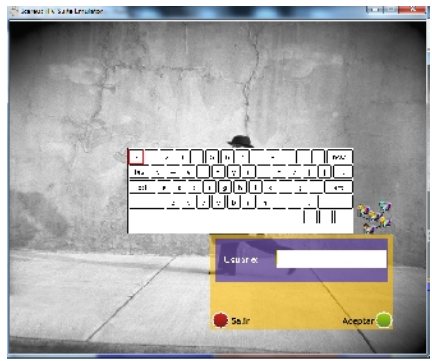

Figure 5. User interface - Virtual keyboard

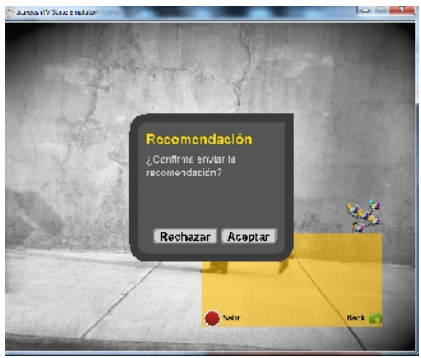

Figure 6. User interface – Dialog

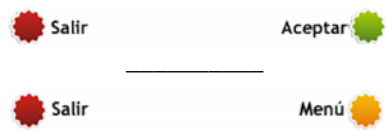

Figure 7. Examples of control bars

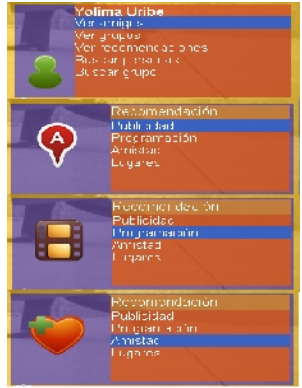

Figure 8. User interface - preview of each option

### 2.4.2.2. External components

The external components are developed in an individual and personalized way according to the needs of each application. The use of these components offers big benefits to the user through the application handling, improve it with specific actions that make it adaptable to the required behavior. These external components are development in a programming language and later they are integrated to the user interface.

The main external components, according to the system requirements of the study case (publicity recommendations), are:

- Principal: This has to give the access and load the user data (profile and photo).
- See friends: This has to load the data (profile and photo) of the current user's friends.
- See groups: This has to load the data (profile and photo) of the current user's groups.
- Search a person: This has to do the search of a person by his name.
- Search group: This has to do the search of a group by his name or topic.
- Do recommendation: This has to list the user's publicity recommendations, confirm the user whom is sending the recommendation (which have to be friend of the current user) and send the recommendation.
- Recommendation: This has to load the recommendation information (Title, content).

With this principal functions list, the next step was their development. To do that, it was necessary to count with others tools dedicated to the development.

1. For the development of external components: Eclipse IDE (development in java language)[1] with Osmosys Plugin.
2. For web services development: because of the styles of architecture, the separation of the composition, the design and logic of an app, because it is an app client-server and the service separation, it was necessary to do the processing through of web services. In his way, that app or the access device (which usually doesn't have good processing capabilities) don't do the hard part of processing of the system, so its job is request and receive the needed information. For develop them, it was necessary to have Eclipse IDE (development in java language) with Axis2[2] Plugin and Apache Axis2[3] server.

---

[1] http://www.eclipse.org/
[2] http://axis.apache.org/axis2/java/core/tools/index.html
[3] http://axis.apache.org/axis2/java/core/docs/installationguide.html

Computer Science & Engineering: An International Journal (CSEIJ), Vol.2, No.6, December 20123. For the data base development: to design and deploy the used data base, we used MySQL Workbench 5.2[4] and a MySQL 5.5.16[5] server.

Working with these tools, it is possible to develop the external components and request a web service which uses the data base. Then, each component in Icareus iTV Suite 3.9 was configured in order to use them inside the application.

### 2.5. Integration

Each external component is responsible for correlating profile data and those recollected from the interaction among users into the social network. The component related with the recommendations, help to reinforce this correlation with the association rules defined like method to do the recommendations (see Table 3).

Table 3. Association rules

| Terms | | Result | Terms | | Result |
| User profile | | Recommendation | User profile | | Recommendation |
| Gender | Age | Type | Age | Activity preference | Type |
| --- | --- | --- | --- | --- | --- |
| 0 | 0 a 10 | 17 | 7 a 40 | 3 | 5 |
| 0 | 11 a 30 | 18 | 7 a 40 | 0 | 5 |
| 1 | 19 a 60 | 6 | 7 a 40 | 1 | 5 |
| 0 | 0 a 3 | 22 | 7 a 40 | 0 | 7 |
| 0 | 4 a 10 | 23 | 7 a 40 | 1 | 7 |
| 0 | 11 a 18 | 25 | 7 a 10 | | 1 |
| 0 | >19 | 26 | 11 a 18 | | 2 |
| 1 | 0 a 3 | 22 | 19 a 40 | | 3 |
| 1 | 4 a 10 | 24 | 19 a 50 | | 4 |
| 1 | 11 a 18 | 25 | 7 a 50 | | 21 |
| 1 | >19 | 27 | 11 a 50 | | 16 |
| | | | >18 | | 8 |
| | | | >18 | | 9 |

---

[4] http://dev.mysql.com/downloads/workbench/
[5] http://dev.mysql.com/downloads/mysql/

25

In these association rules, the recommendation type can be: 1: Academic-elementary, 2: Academic-high school, 3: Academic-undergraduate, 4: Academic-Postgraduate, 5: Sport, 6: Esthetics, 7: Movies, 8: News, 9: Information, 10: Food, 11: Music, 12: Religion, 13: Adults, 14: Health, 15: Home, 16: Technology, 17: Child games, 18: Teenage games, 19: Culture, 20: Events, 21: General academic, 22: Babies clothing, 23: Boys clothing, 24: Girls clothing, 25: Teenagers clothing, 26: Men clothing, and 27: Women clothing.

Having the external components deployed into user interface, the next step was to set the complete app on the television server. To do this, we integrated the app to the application server using Icareus iTV Playout, tool which manages the content of the TV server through channels and its content. For this, it is necessary to have the app finished, the each external components configuration files and set them into the server and deploy them on a display channel. After this process, the application can be deployed and accessed from a decoder connected to apps server.

## 2.6. Validation

As the developed application is a part of practice validation of the architectural designed mentioned, its implementation has the goal of verify the adaptation model, specially implementing user profile, group profile and social profile, the social net model and the architectural styles: *i)* Client- server, *ii)* Architecture by layers, *iii)* Separation of the composition, design and logic of an application, *iv)* Sessions management, *v)* Services separation, *vi)* Management, publication and distribution of information, *vii)* Management of resources information and *viii)* Light client. The results are presented below.

### 2.6.1. Results

The final client is a small app with the minimal resources needed to execute it, complying with the "light client" required.

The final app is made according to the layers of the architecture and its performance is correct in respect of the main models.

The adaptation and social network models were applied with data of the mentioned three profiles and the interaction layer was made under the association rules developed according to the relationships between the designed profiles. These recommendations were applied and transmitted from user to user, according to the defined rules of social network. This shows that the aspects to validate with this study case were well in implementation and in deployment, for this reason, we hope a good performance of the app with all aspects of the architecture.

Between the aspects to have in account in the implementation of this type of apps, are the tools to develop the external components. In this case particularly, one of the biggest issues was the use and dependence of the Axis2[6] classes for the web services request inside of an external component, because of the final application size. Although, it didn't exceed the size supported for the server, the needed classes to represent an additional size for the application, which must be considered before use some tool in the development. This detail wasn't taken in account in the beginning of the process because of the lack of the use and performance of this kind of tools to digital television.

---

[6] http://axis.apache.org/axis2/java/core/

## 3. RELATED WORK

There have been apps about the use of: digital television, service adaptation and social network together, but those works and apps have been treated in a separate way or using some of these areas in an app, but there is not one that meets all the three topics. For this reason, we expose the related works with this project independently.

In first place, taking the issue of user interaction-TV, we found two interesting works: "Spontaneous interaction with audiovisual contents for personalized e-commerce over Digital TV" [18] and "The Digital Television Multimedia Message Service Flows" [19], where is important highlight their role in the context in which the applications are implemented, because their focus is the user and not the machine or the tool itself. Some important developments are the methods, platforms and tools used for each implementation, because these apps are so technical and offer a framework of this point.

Related to the user – user interaction in digital television, we found some applications which are focused in offering extra services for digital TV like "A T-Learning Platform based on Digital Terrestrial Television" [20], "Spontaneous interaction with audiovisual contents for personalized e-commerce over Digital TV"[18], "Exploiting digital TV users' preferences in a tourism recommender system based on semantic reasoning" [21] and "Proposed model of a digital video-based home telecare system" [22]. These works show how the applications and tools that are different to the TV content can be supported in this network. But, taking advantage of this network and of the contents that are handled in it, the services can be enriched with such information. Also, how the interaction is among users, the preferences of each one represent an important difference to make the user feels comfortable and participant in the application.
In second place, we found some apps related to social networks and recommendation systems, like a eco tourism system [23] and a recommendation system based on a social network [24], where the main work there, is to use the social structure data to offer to users benefits according to some user's needs.

Other interesting applications are those which work as interaction facilitator. As mentioned above, the center of the social net is the user, so, the relationships between them have a high priority, because of indeed, each social activity is guided for the use of the social network [25].

## 4. CONCLUSIONS

- Using the interaction among users as base to offer services, allow to give better benefits to users, because these services are going to be according to the information that we can extract of the interactions, like the involved users, the type of correlation, the subject and the related services, and to use this data for adapt services. In this way, it supports the diffusion techniques and gives to users a better experience inside a system.
- The use of adaptation as tool to enrich services, allows to system to be prepared to face different types of data, information, access devices, users and them preferences, so the use environment for a client will be different for each one, compliance with specific aspects of each like a personal preference or liking.
- The development of a digital TV application along with a social network and with adapted services is a new proposal which offers benefits for developers, users and providers. Also, it opens doors for the use, for research and for the implementation of such tools to open the market and the possibilities for this kind of applications.

## ACKNOWLEDGEMENTS

To the Pontificia Universidad Javeriana – Bogotá – Colombia, especially to the program of Master in Information Systems and Computer Engineering and its research group SIDRe (Information systems, distributed systems and networks).

## Authors

Mery Yolima Uribe Rios MSc is a systems and computing engineer and magister, how works and researches in networks, agents systems, digital television, security and grid computing.

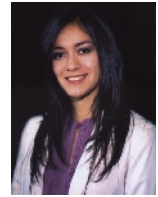

Rafael V. Páez PhD. works at Pontificia Universidad Javeriana and is one of the head of in the research group SIDRe in distributed systems and networks area.

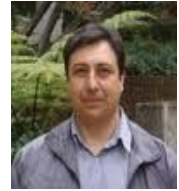